\documentclass[%
 reprint,
 aip,
 amsmath,amssymb,
twocolumn,
groupedaddress
]{revtex4-1}

\usepackage{graphicx}
\usepackage{dcolumn}
\usepackage{bm}

\usepackage[utf8]{inputenc}
\usepackage[T1]{fontenc}
\usepackage{mathptmx}
\usepackage{braket}
\usepackage{xcolor}

\begin{document}


\title{Accelerating quantum optics experiments with statistical learning}

\author{Cristian L. Cortes}
\author{Sushovit Adhikari}
\author{Xuedan Ma}
\author{Stephen K. Gray}
\affiliation{Center for Nanoscale Materials,\\ Argonne National Laboratory,\\ Lemont, Illinois 60439, USA}

\date{\today}

\begin{abstract}
Quantum optics experiments, involving the measurement of low-probability photon events, are known to be extremely time-consuming. We present a new methodology for accelerating such experiments \textcolor{black}{using physically-motivated ansatzes together with simple statistical learning techniques such as Bayesian maximum a posteriori estimation based on few-shot data.} We show that it is possible to reconstruct time-dependent data using a small number of detected photons, allowing for fast estimates in under a minute and providing a one-to-two order of magnitude speed up in data acquisition time. We test our approach using real experimental data to retrieve the second order intensity correlation function,
$G^{(2)}(\tau)$, as a function of time delay $\tau$ between detector counts, for thermal light as well as  anti-bunched light emitted by a quantum dot driven by periodic laser pulses. The proposed methodology has a wide range of applicability and has the potential to impact the scientific discovery process across a multitude of domains.
\end{abstract}

\maketitle
\newpage
Intensity interferometry is a hallmark technique in quantum optics for determining the statistical properties of light based on correlated photon events \cite{glauber1963quantum,boal1990intensity,baym1998physics,dravins2013optical}. In the original Hanbury Brown and Twiss (HBT) proposal \cite{brown1958interferometry}, two detectors were used to perform continuous, steady-state measurements of photons with respect to detection time and separation distance, \textcolor{black}{allowing for the characterization of stars.}  \textcolor{black}{Nowadays, this technique is used} to characterize and categorize light sources with applications in quantum computation, communications, metrology \cite{yuan2002electrically,mckeever2004deterministic,pelton2002efficient,kuhn2002deterministic,michler2000quantum,waks2002secure,hogele2008photon,grosse2007measuring,schulte2015quadrature,carmele2009photon,brod2019photonic,kaneda2019high,wang2018experimental,loredo2017boson,cooper2013experimental}, and imaging \cite{lugiato2002quantum,brida2010experimental,tsang2009quantum,tenne2019super,forbes2019super,classen2017superresolution}. \textcolor{black}{In practice}, non-idealities, such as imperfect experimental conditions or intrinsically weak light sources, often make the probability of detecting two-photon events extremely low. Acquisition times can range from seconds, to minutes, to hours \cite{girish2017high,schneeloch2018quantifying}, and sometimes days \cite{abbott2016characterization}, making the technique limited and unappealing in many applications, most notably in situations where raster scanning is required. Here, \textcolor{black}{we show that physically-motivated ansatzes together with} statistical learning techniques, such as Bayesian maximum a posteriori estimation, provide several order of magnitude speed-ups in parameter estimation. 

In this Letter, we focus on the reconstruction of the second-order intensity correlation function, $G^{(2)}(\tau)$, under stationary and non-stationary conditions. The proposed methodology shows that emerging concepts in the fields of statistical and machine learning could have a tremendous impact in accelerating scientific exploration in the quantum optical domain. While machine learning approaches are well known for image reconstruction and denoising under low light conditions \cite{gull1978image,kim1990recursive,morris2015imaging}, and very recently with interesting work on the classification of light sources \cite{kudyshev2019rapid,you2019identification}, there is much less work within the framework of time-dependent phenomena in quantum optics. Tackling such problems is important for reducing the total time required to perform experiments as well as {enabling} experiments which would be hard to perform otherwise due to the weak photostability of molecules or a low signal-to-noise ratio. 


\begin{figure}
    \centering
    \includegraphics[width=8.1cm]{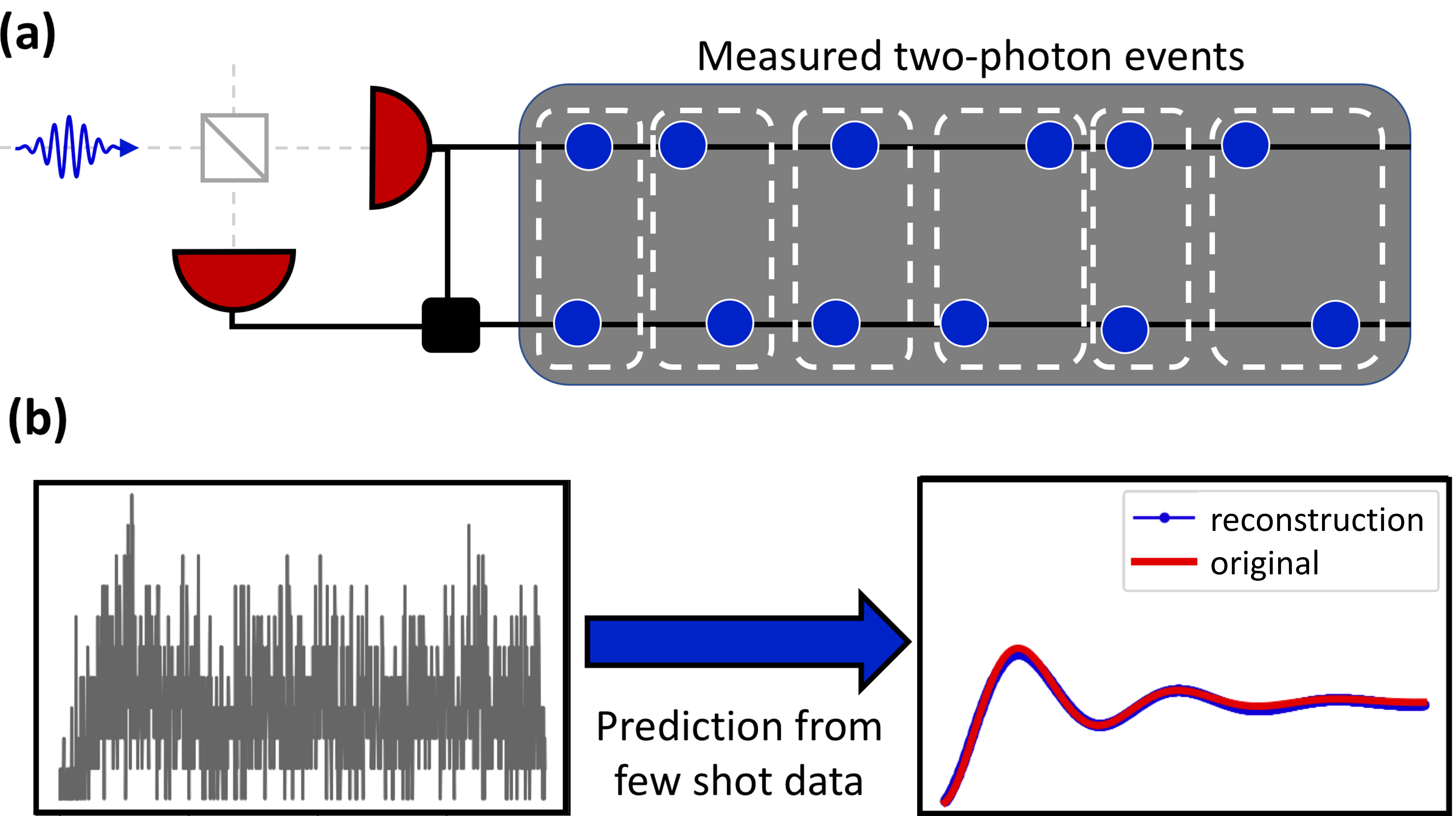}
    \caption{(a) Quantum optical experiments, such as HBT interferometry, rely on the detection of low-probability photon events. (b) Using only hundreds of photons, less than one per time bin on average, we show that simple statistical learning methods are able to reliably reconstruct the true signal from few-shot data. This can provide a tremendous speed-up in the data acquisition process for a wide variety of applications. The two-photon events shown in the top panel are representative of bunched light emitted from a thermal light source, while the signal shown in the bottom is representative of anti-bunched light emitted from a single quantum dot. }
    \label{fig:my_label}
    \vspace{-0.5cm}
\end{figure}
\begin{figure*}[t!]
    \centering
    \includegraphics[width=16cm]{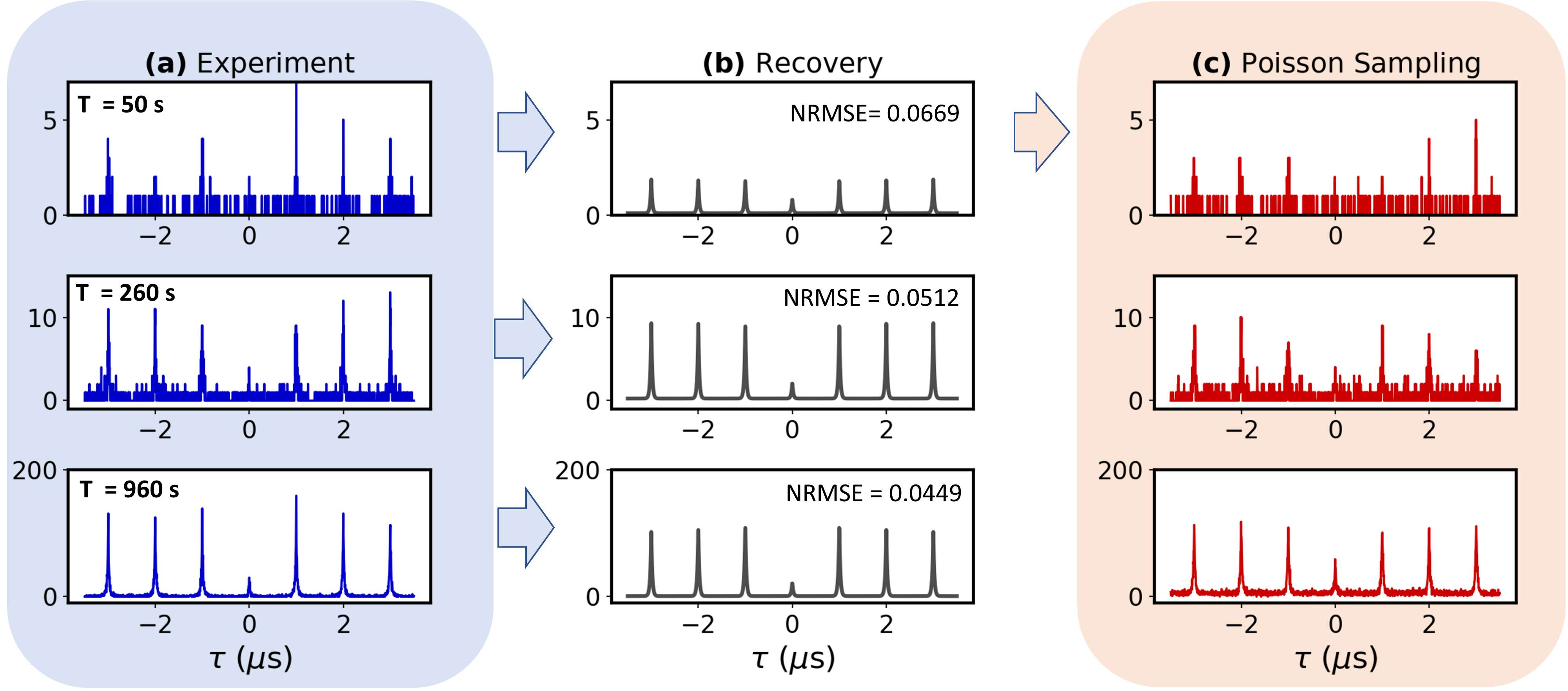}
    \caption{Second-order photon correlation spectroscopy from a single photon emitter. A single quantum dot is driven by a periodic train of picosecond laser pulses. (a) The count-based second-order intensity correlation function is shown for total integration times of $T$ = 50 s, 260 s, 960 s. (b) The noiseless reconstructed signal is shown for each case using Bayesian MAP estimation\textcolor{black}{, together with the normalized root-mean-squared error for the recovered signal compared to the raw 960 s experimental result}. (c) Poisson sampling simulations of the 50 s recovered signal are shown for various simulated integration times, showing excellent agreement between the algorithm's predictions and true experimental results. \color{black}{The background contribution from the correlation function is subtracted in all plots resulting in a drop to zero in the long time limit.}}
    \label{fig:my_label}
    \vspace{-0.5cm}
\end{figure*}

The \textcolor{black}{focus of this work is the} second order intensity correlation function~\cite{scully1999quantum},  $G^{(2)}(\tau)$ = $\braket{:\hat{n}(t)\hat{n}(t+\tau):}$, with $\hat{n}(t)$ being the photon number operator at time $t$,  which is proportionate to the joint probability of detecting a photon at $t$ on a start detector and detecting a second photon at $t+\tau$ on a stop detector (see Fig. 1). The brackets correspond to an average over $t$. It should be noted that these ideas are  applicable to a wide variety of quantum optics settings and other types of quantum measurements that are based on sampling. 

\paragraph*{Bayesian maximum a posteriori estimation.}
Our approach is based on Bayesian statistical modeling. We aim to find an underlying time signal, $\mathbf{y}=(y_1,y_2,y_3,\cdots,y_M)$, 
given an \emph{incomplete} set of measurements $\mathbf{n} = (n_1,n_2,n_3,\cdots,n_M)$ with each $n_i$ being the number of photons observed in time bin $i$. The quantity $\mathbf{y}$ refers to the number of two-photon events in specific time bins and is used to construct the second-order correlation function, $G^{(2)}(\tau)$. 
Signal reconstruction based on few-shot data, as shown in Fig. 1, naturally fits within the framework of Bayesian inference. A straightforward implementation of this framework is known as \emph{maximum a posteriori} (MAP) estimation \cite{kay1993fundamentals} which aims to maximize the logarithm of the posterior probability $p(\mathbf{y}|\mathbf{n})$. Using Bayes' theorem, the posterior distribution may be written as, $p(\mathbf{y}|\mathbf{n}) = p(\mathbf{n}|\mathbf{y}) p(\mathbf{y})/p(\mathbf{n})$, where $p(\mathbf{n}|\mathbf{y})$ is defined as the likelihood which, at the few-photon level, is equal to the product of Poisson probability distributions, $p(\mathbf{n}|\mathbf{y}) = \prod_i y_i^{n_i} e^{-y_i}/(n_i!).$ 
Each two-photon event in a given time bin, which represents a measured shot, is assumed to be a statistically independent event. The Poisson distribution then describes the probability of detecting $n_i$ photons given the expected value $y_i$. 
$p(\mathbf{y})$ describes the prior knowledge of the underlying time signal, and $p(\mathbf{n})$ describes the marginal likelihood which acts as a normalization factor.
The use of Poisson distributions to describe the experimental shot statistics should not be confused with any assumption about the photon statistics of an emitter being studied. The
emitter need not exhibit Poissonian statistics and, indeed, the single-photon emitter
example to be studied later is one with anti-bunching and sub-Poissonian statistics.    

In typical signal reconstruction algorithms, optimization proceeds with the maximization of the log-likelihood with $\mathbf{y}$ acting as the set of tunable parameters. For large data sets, this becomes a time-consuming optimization problem. To obtain fast estimates, we instead parameterize the signal, $\mathbf{y}\rightarrow \mathbf{y}(\boldsymbol{\theta})$, where $\boldsymbol{\theta} = (\theta_1,\theta_2,\theta_3,\cdots,\theta_N)$ becomes the new set of optimization parameters. By maintaining the number of parameters small, it becomes easier to obtain good estimates using only a few shots. Bayesian maximum a posteriori estimation proceeds with maximizing the logarithm of the posterior because it simplifies the objective function while preserving the maxima. The optimization objective is then given by, $\max_{\boldsymbol{\theta}}\; \log p(\mathbf{y}(\boldsymbol{\theta})|\mathbf{n}) =\log p(\mathbf{n}| \mathbf{y}(\boldsymbol{\theta})) + \log p( \mathbf{y}(\boldsymbol{\theta}))$, which may be written explicitly as:
\begin{align}
   \max_{\boldsymbol\theta}\; \sum_i ( n_i\log y_i(\boldsymbol{\theta}) - y_i(\boldsymbol{\theta})  ) - \sum_j \lambda_j |\theta_j|,
\end{align}
where $\lambda_j$ is a hyperparameter related to the chosen distribution function of the prior. See supplementary material for more detail.  The prior may be interpreted as a regularization term in the objective function, which may be physically motivated for a wide variety of systems. \textcolor{black}{Equation (1) results from taking the prior to be a Laplace distribution, 
$p( \mathbf{y}(\boldsymbol{\theta})) = \prod_j \tfrac{1}{2\sigma_j} e^{-|\theta_j|/\sigma_j}$ on a chosen subset of parameters $\theta_j \in \{\boldsymbol{\theta}\}$. For example, the subset could correspond to the amplitudes of damped sinusoids. In the most general case, each parameter $\theta_j$ has a different standard deviation $\sigma_j$ = $\lambda_j^{-1}$. One can relax this assumption and take them to all have the same standard deviation. This yields a single hyperparameter, $\lambda$, which is often encountered in optimization problems with Lasso regularization.\cite{Candes2008} Such a regularization term is related to sparsity and is used in compressed sensing for signal and imaging reconstruction.\cite{Candes2008} Since we expand the second-order photon correlation function in terms of a series of damped sinusoids, Lasso regularization can  help in finding a suitable fitting function with only a small number of terms.} If we ignore the prior by setting $\lambda = 0$, which is equivalent to assuming a uniform distribution for the estimation parameters, this procedure reduces to \emph{maximum likelihood estimation}. In certain cases, the maximum likelihood estimation is sufficient for obtaining fast estimates, as we show below.

Finding an optimal set of parameters $\boldsymbol{\theta}$ can be done using a wide variety of approaches. If the parameterization is chosen to lie within the family of exponential functions, the objective function will be convex allowing optimization to proceed using a wide-variety of convex-optimization approaches \cite{roux2012stochastic,johnson2013accelerating,bubeck2015convex}. \textcolor{black}{When the parameterization is more complex, as is the case in this manuscript, the objective might not be convex requiring careful optimization procedures commonly encountered in deep learning with neural networks.} We tested a wide variety of different optimization subroutines but found that Powell's conjugate direction  method \cite{powell1964efficient} is consistently the most successful in finding near optimal solutions in a short amount of time. Furthermore, we found that multi-start optimization with several initial guesses is required to ensure the best solution is found. In principle, the guesses are independent of each other, therefore, this step is highly parallelizable and can be performed quickly with multi-threading or GPU computing.

\begin{figure*}[t!]
    \centering
    \includegraphics[width=16cm]{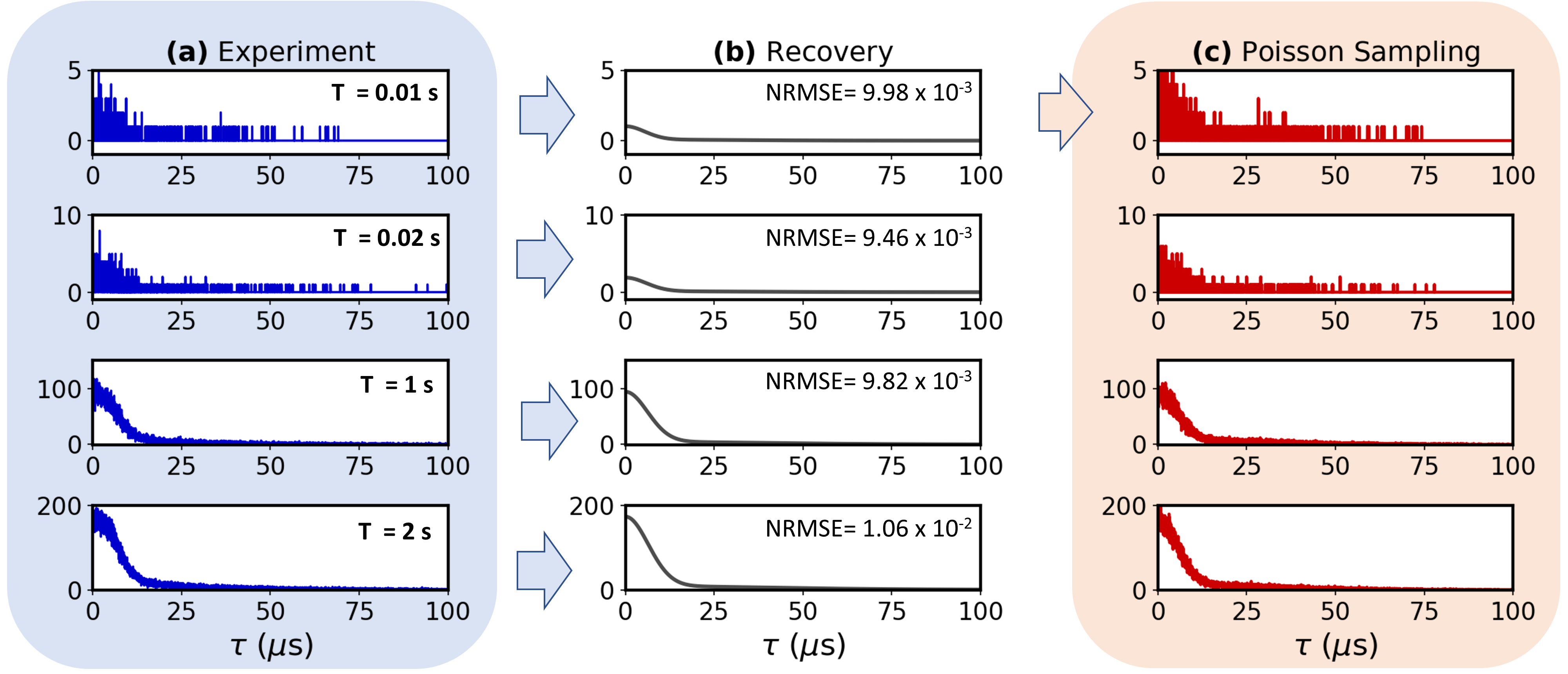}
    \caption{Continuous, steady-state thermal emission from a neon light source. (a) Measured second-order correlation function $G^{(2)}(\tau) $ using raw counts for integration times of $T$ = 0.01s , 0.02 s, 1,2 s. (b) Reconstructed noiseless signal for each case\textcolor{black}{, together with the normalized root-mean-squared error for the recovered signal compared to the $T$ = 2 s experimental result}. (c) Poisson sampling simulation of 0.01 s reconstruction showing excellent agreement with experimental results in (a). \color{black}{The background contribution from the correlation function is subtracted in all plots resulting in a drop to zero in the long time limit.}}
    \label{fig:my_label}
    \vspace{-0.5cm}
\end{figure*}

We now test the proposed methodology with real experimental data. First, we consider measurements of the second-order intensity correlation, $G^{(2)}(\tau)$, of a single CdSe/CdS core/shell quantum dot driven by periodic picosecond laser pulses, shown in Fig. 2. The experimental results are shown in the left column for integration times of 50, 260, and 960 seconds respectively. In a typically measurement, emission from a single quantum dot is collected by a microscope objective, separated by a 50/50 beam splitter, and sent to two identical single photon detectors. The correlation events between the two photon detectors are recorded as a function of the delay time. As is commonly observed in quantum emitters, local field fluctuations can induce photoluminescence blinking, which could severely prolong the integration time of second-order photon correlation measurements \cite{efros2016origin,ma2010fluorescence}. For this time-dependent case, we derive the following parameterization for the signal:
\begin{equation}
  y_i(\boldsymbol{\theta}) = c_0 +  c_1 e^{- \gamma_1 |\tau_i |}\left( c_2 e^{-\gamma_2|\tau_i|} +\sum_{n\neq 0}  e^{-\gamma_2|\tau_i - n \Lambda |}  \right) ,
\end{equation}
where $\boldsymbol{\theta} = \{c_0,c_1,c_2,\gamma_1,\gamma_2,\Lambda \}$ and $c_0$ is a background signal that is dependent on dark current as well as other background noise (see supplementary material for details). The algorithm is faster if the background is characterized beforehand, however, it is possible to leave $c_o$ as a free parameter. \textcolor{black}{ The parameter $c_1$ is an amplitude for the correlation function that affects every peak, while $c_2$ is a multiplicative factor that affects the height of the $G^{(2)}(0)$ peak only. $\Lambda$ is the repetition period for the pulsed laser. $\gamma_1$ and $\gamma_2$ are two different decay rate constants that are necessary to describe three-level systems; in general, it is possible to generalize the result to have multiple decay rates, however, we find that this result is sufficient for the current work. } We use the short 50s trial as the input, yielding a noiseless signal estimate,  $y(\boldsymbol{\theta})$ (middle column, top row). For reference, we also show the estimates from the 260s and 960s trials (middle column, bottom two rows).  \textcolor{black}{Each recovered signal includes the normalized root-mean-squared-error (NRMSE) defined with respect to the long acquisition time signal (bottom left plot). Normalization is relative to the difference between maximum and minimum signal values; see supplemental material. The NRMSE shows a decreasing trend as the total number of photons increases}. The general features show excellent agreement apart from differences in the overall amplitude, which are simply attributed to optimization being performed on the samples with different numbers of photons. 

To exemplify how well the noiseless estimate from the 50s input performs, we perform Poisson sampling simulations shown in the third column of Fig. 2; see supplementary material for details. We use $y(\boldsymbol{\theta})$ from the 50s estimate only, but include a multiplicative factor $T$, $y\rightarrow y\times T$, representing the integration time in the experiment. We chose the integration time to approximately match the total number of photons detected for the 260s and 960s cases respectively. Comparing the first and third columns illustrates the performance of the signal extraction technique in predicting experimental results for longer integration times. In particular, the predicted lifetime of the emitter show excellent agreement. While the zero-time second-order correlation $G^{(2)}(\tau=0)$ is over estimated compared to the true result, the 50s result is still able to predict anti-bunching, indicating the quantum dot's potential as a single-photon source. 

Next, we consider the characterization of a neon discharge lamp acting as a thermal light source under continuous steady-state conditions. The experimental results for various integration times are shown in the left column of Fig. 3 with noiseless signal estimates $y(\boldsymbol{\theta})$ shown for each case in the second column. In this experiment, we use a parameterization based on the sum of Gaussians:
\begin{equation}
  y_i(\boldsymbol{\theta}) = c_0 + \sum_{n=1} c_n\exp{ (-\frac{\tau_i^2}{2\sigma_n^2}) }
\end{equation}
where $\boldsymbol{\theta} = \{c_0,c_1,\sigma_n \}$. This expression is known to describe inhomogeneously broadened thermal light sources (see supplementary material). Once again, we use the short integration time result ($0.01 s$) as the input to perform simulated Poisson sampling experiments (third column), which yields simulated experimental estimates of the second-order correlation function for various integration times $T$. In all cases, we find excellent agreement between the experimental results as well as the algorithm's predictions, \textcolor{black}{quantified by the small normalized RMSE. Unlike Figure (2), the normalized RMSE for all four integration times are relatively similar within sampling error. This can be expected because, in general, signal recovery will be more error-prone for small number of photons but will become relatively stable once a certain number of photons have been collected, which happens to be the case for the current experiment.} We note that this result provides a two order of magnitude speed-up in the data acquisition process.

\paragraph*{Precision of measurements.} We now discuss limits on the precision of the signal recovery procedure outlined in this manuscript. The Cramer-Rao bound \cite{kay1993fundamentals} provides a lower bound on the variance of the unbiased estimate of a signal $y_i$, $\text{var}(\hat{y}_i) \geq (F^{-1})_{ii} = y_i,$
where $F$ is an $M\times M$ matrix representing the Fisher information, $M$ being the
total number of time bins. This bound implies that the variance will be strictly greater than, or equal to, the \emph{mean} of the signal, as expected due the nature of Poisson statistics -- the equality is proved in the supplementary material. This implies that, at best, the precision is shot-noise limited when using unbiased estimation techniques. However, it is well known that image denoising algorithms provide \emph{biased} estimation \cite{chatterjee2009denoising,levin2011natural}, therefore the variance adheres to the generalized Cramer-Rao lower bound, allowing for more precise measurements with the introduction of a bias as is done in the present manuscript.
We now discuss possible applications for this methodology, as well as possible extensions to this approach using deep learning.

\paragraph*{Characterization of quantum light sources.} Quantum light sources are valuable for enabling emerging quantum information technologies \cite{hogele2008photon,grosse2007measuring,schulte2015quadrature,carmele2009photon,brod2019photonic,kaneda2019high,wang2018experimental,loredo2017boson,cooper2013experimental}. For example, light sources that generate on-demand, indistinguishable single photon Fock states are important for boson sampling \cite{aaronson2011computational,lund2014boson,hamilton2017gaussian} and quantum computation/communication \cite{gisin2007quantum,tanzilli2005photonic,o2009photonic}. The generation and characterization of multi-photon cluster states are also important for one-way quantum computing \cite{nielsen2004optical,chen2007experimental}. Finding and characterizing emitters that generate these states of light from quantum dots, defects in diamond or 2D materials represents a time-consuming step. Accelerating this step would provide at least an order-of-magnitude speed up in the characterization process.

In biological imaging, fluorophores are  used as markers that can be detected through the emission of photons. Often, these fluorophores are metastable under continuous pumping, transitioning into a trap state where they cannot emit photons \cite{ma2010fluorescence}. Under these conditions, it is desirable to obtain clear estimates of the fluorophore's properties for short integration times. Our methodology may impact this field, allowing the processing of data previously thought to be too noisy for signal extraction.

\paragraph*{Quantum imaging.}
We envision that our methodology will have the greatest impact in quantum super-resolution imaging \cite{lugiato2002quantum,brida2010experimental,tsang2009quantum,tenne2019super,forbes2019super,classen2017superresolution}. Quantum imaging uses two-photon counts to perform image reconstruction beyond the diffraction limit. The second-order correlation signal provides a factor of $\sqrt{2}$ improvement, while $n$-order correlation signals  provide $\sqrt{n}$-times improvement in the resolution. Since higher-order signals becomes less and less probable, the integration times for obtaining sub-diffraction resolution becomes prohibitively long. Furthermore, even in the two-photon confocal microscopy using raster scanning, this approach is  known to be time consuming. Our proposed methodology would be able to provide dramatic speed-ups paving the way for real-time sub-diffraction imaging in the near future. 


To conclude, we discuss the performance of our approach compared to other numerical methods. The most widely used approach for curve fitting is the least squares method, typically using Levenberg-Marquardt optimization \cite{marquardt1963algorithm}. In the supplementary material, we provide a thorough analysis of how this method compares to the MAP technique outlined here. Generally, we find that the least squares estimate has much higher variance than the Bayesian approach using the objective function Eq.(2). This implies that from sample to sample, the least square estimate can provide wrong estimates more often than not. However, the least squares method surprisingly does a good job for a wide variety of cases when the multi-start approach is included. In general, we found the Bayesian MAP approach did just as well or outperformed the least squares method, allowing us to conclude that this should be the preferred approach. 
We emphasize that both the least squares method, as well as the Bayesian MAP or maximum likelihood approaches, provide an important benchmark for signal reconstruction for which all future methods should be compared to. Deep learning approaches using convolutional neural networks for supervised machine learning \cite{jain2009natural,vincent2008extracting,kudyshev2019rapid}, or autoencoders in unsupervised machine learning \cite{vincent2008extracting,vincent2010stacked,chen2012marginalized}, will have to provide significantly better improvements to the results of the present manuscript. Furthermore, the black-box nature of the neural network approaches will always lack the transparency of these simple statistical learning approaches, therefore, they might always be less appealing for certain applications. We anticipate that there will be rapid improvement in low-photon signal reconstruction in the next few years.


\paragraph*{Acknowledgements.}
This material is based upon work supported by Laboratory Directed Research and Development (LDRD) funding from Argonne National Laboratory, provided by the Director, Office of Science, of the U.S. Department of Energy under Contract No. DE-AC02-06CH11357. 
Use of the Center for Nanoscale Materials, an Office of Science user facility, was supported by the U.S. Department of Energy, Office of Science, Office of Basic Energy Sciences, under Contract No. DE-AC02-06CH11357.

\bibliographystyle{apsrev}
\bibliography{AccQO.bib}

\end{document}